\documentclass[aip,reprint,preprintnumbers]{revtex4-1}
 
\usepackage{color}
\usepackage{siunitx}
\usepackage{graphicx}
\usepackage{amsmath}
\usepackage{hyperref}
\usepackage{xurl}
\usepackage{bm}

\makeatletter
\def\@email#1#2{%
\endgroup
\patchcmd{\titleblock@produce}
{\frontmatter@RRAPformat}
{\frontmatter@RRAPformat{\produce@RRAP{*#1\href{mailto:#2}{#2}}}\frontmatter@RRAPformat}
{}{}
}%
\makeatother

\begin{document}


\preprint{\parbox[c][1cm][c]{\textwidth}{
This article may be downloaded for personal use only. Any other use requires prior permission of the author and AIP Publishing. This article appeared in J. Chem. Phys. \textbf{165}, 131103 (2026) and may be found at \url{https://doi.org/10.1063/5.0343876}.}}

\title{Hydrodynamics substantially affects induced structure formation in magnetic fluids}

\author{Henning Reinken}
\affiliation{Otto-von-Guericke-Universität Magdeburg, Institut f\"ur Physik, Universitätsplatz 2, 39106 Magdeburg, Germany} 

\author{Markus Heiber}
\affiliation{Otto-von-Guericke-Universität Magdeburg, Institut f\"ur Physik, Universitätsplatz 2, 39106 Magdeburg, Germany}  

\author{Takeaki Araki}
\affiliation{Kyoto University, Department of Physics, Kyoto 606-8502, Japan} 

\author{Andreas M. Menzel}
\email{henning.reinken@ovgu.de, a.menzel@ovgu.de}
\affiliation{Otto-von-Guericke-Universität Magdeburg, Institut f\"ur Physik, Universitätsplatz 2, 39106 Magdeburg, Germany}

\date{\today}

\begin{abstract}
Magnetorheological fluids consist of micrometer-sized magnetic particles in a carrier
liquid. Sufficiently strong external magnetic fields lead to the formation of string-like particle aggregates. 
We demonstrate that hydrodynamic interactions, that is, mutual couplings via induced flows, play a substantial role during the structuring process. 
They support the formation of slender chains instead of more compact clusters in the absence of mutual hydrodynamic interactions between the particles.
This fundamental insight is substantial from an application perspective, due to the enormous technical importance and potential of structured magnetorheological materials.
\end{abstract}


\maketitle 

For decades, the appealing properties of magnetic fluids have been explored. Stabilizing nanometer-sized magnetic particles in carrier liquids leads to magnetic suspensions. Such ferrofluids~\cite{rosensweig2013ferrohydrodynamics} are of outstanding technical relevance, considering, for instance, their use in hard drives or loudspeaker components. Suspensions of larger, micrometer-sized magnetic or magnetizable particles are referred to as magnetorheological fluids~\cite{bossis2002magnetorheological,vicente2011magnetorheological}. The name reflects their remarkable feature of tunable shear viscosity. More precisely, the magnitude of the overall viscosity can be substantially affected by application of a sufficiently strong, homogeneous, external magnetic field~\cite{bossis2002magnetorheological,vicente2011magnetorheological}. 

The background of this so-called magnetoviscous effect had been discussed for a longer while. Its root was searched for in rotational blocking of individual particles by external magnetic fields~\cite{shliomis1972effective}, specifically for ferromagnetic particles.  
Yet, this effect is often not sufficient to explain the observed substantial change in overall viscosity in magnetorheological fluids. Instead, the magnetically induced formation of string-like aggregates plays a central role in revealing the source of the magnetoviscous effect~\cite{carlson2000mr,vicente2011magnetorheological}. 
When an external magnetic field is applied, the magnetized particles assemble into anisotropic string-like aggregates that favor orientation along the external magnetic field.
These oriented structures resist reorientation under shear and thus lead to a significantly enhanced apparent viscosity of the suspension~\cite{vicente2011magnetorheological}.
Thus, the phenomenon is based on collective processes of microscopic, particulate structure formation.

There are further circumstances where the formation of string-like aggregates of magnetic particles becomes important. For instance, magnetorheological elastomers consist of micrometer-sized magnetic or magnetizable particles fixated in an elastic carrier medium~\cite{filipcsei2007magnetic,behrens2011preparation,odenbach2016microstructure,bastola2020recent}. 
These materials likewise show a magnetorheological effect. Imposed, homogeneous, external magnetic fields affect their overall stiffness and damping behavior~\cite{jolly1996model,gong2005fabrication,lokander2003improving,pessot2018tunable}. 
Moreover, they change their shape in external magnetic fields. This magnetostrictive behavior~\cite{stolbov2011modelling,metsch2016numerical,romeis2017theoretical,fischer2019magnetostriction,sanchez2019modeling,silva2022giant,roghani2025magnetically} suggests their use as soft actuators, soft robots, or magnetic valves~\cite{schmidt2006electromagnetic,bose2012soft,fuhrer2013soft,kim2022magnetic}.

It turns out that imprinted structuring of the magnetic particle distribution significantly affects these types of behavior. String-like aggregates can substantially enhance the nature and magnitudes of both magnetorheological and magnetostrictive effects~\cite{martin2006magnetostriction,ivaneyko2012effects,metsch2016numerical,fischer2019magnetostriction,filipcsei2007magnetic,stepanov2007effect,bose2009magnetorheological,ivaneyko2012effects,han2013field,cremer2015tailoring,schumann2017situ,cremer2016superelastic}. 
Such structured magnetorheological elastomers are produced by inducing the formation of anisotropic particle aggregates in still fluid, reactive suspensions by exposure to homogeneous, external magnetic fields~\cite{carlson2000mr,vicente2011magnetorheological,gunther2012x}. 
These structures are locked in when establishing the permanent elastic carrier medium by chemically crosslinking polymers from the surrounding reactive liquid. Thus, during this process, magnetically induced structure formation in initially fluid systems plays a substantial role. 

Various other contexts exist in which string-like aggregates of magnetic particles are formed. One example are living biological microorganisms that use strings of magnetic particles for orientation in the Earth's magnetic field~\cite{faivre2008magnetotactic}. 

Overall, we may conclude that formation of string-like aggregates from microscopic magnetic or magnetizable particles in surrounding fluids is central in nature and technology. 
Its quantitative and theoretical description, including the coupling to the carrier medium, is crucial to understand the influence of the underlying effects and thus to allow for targeted improvement in the future. 

Corresponding studies in terms of computer simulations face significant challenges. 
Couplings between discrete particles and continuous surrounding carrier media represent one of them. Correct boundary conditions on the moving surfaces of the particles must be maintained. Moreover, mutually attracting magnetic or magnetized particles approach each other from the distance until they are basically in contact. Thus, far-field hydrodynamic expansions~\cite{dhont1996introduction} are not sufficient, while finite-element methods face the problem of strongly distorted calculation meshes~\cite{richter2017fluid}.

In most theoretical studies on structure formation in the given topical context, hydrodynamic interactions beyond Stokes drag are not considered~\cite{ghaffari2015review}. 
Only a few previous works have explicitly resolved the emerging hydrodynamic flows and their consequences.
Corresponding strategies include smoothed particle hydrodynamics~\cite{hashemi2016sph,lagger2015influence}, 
immersed boundary methods combined with finite-volume solvers for the hydrodynamic equations~\cite{fernandes2023particle},
lattice-Boltzmann methods~\cite{han2010modelling,han2010three,fu2019simulation}, and
multi-particle collision dynamics~\cite{zablotsky2019field}.
In addition to investigating structure formation, hydrodynamics has also been included in computer simulations of magnetic filaments~\cite{lusebrink2016behavior,mostarac2022rheology} and membranes~\cite{brisbois2022locomotion}, as well as magnetic gels~\cite{novikau2022behaviour,novikau2026rheology}.
Here, lattice-Boltzmann schemes~\cite{mostarac2022rheology,brisbois2022locomotion,novikau2022behaviour,novikau2026rheology} and multi-particle collision dynamics~\cite{lusebrink2016behavior} have been employed as well.
These approaches determine the dynamics of the discrete magnetic particles via Newton's second law and couple them to the flow of the surrounding continuous carrier medium, which is calculated in parallel. They can become quite involved to ensure the correct boundary conditions between particles and fluid~\cite{fernandes2023particle,hashemi2016sph,han2010modelling,han2010three,fu2019simulation}.
 
A further challenge is the accurate description of magnetic interactions between particles.
We focus on magnetorheological fluids comprising magnetically soft particles of a few or tens of micrometers in size or larger, which are magnetized by an external magnetic field.
Thermal fluctuations lose their significance for such particle sizes~\cite{vicente2011magnetorheological}.
When particles approach each other towards contact, they may mutually reinforce their magnetization~\cite{metsch2016numerical,ghaffari2015review}. 
Magnetization below saturation generally becomes nonuniform within the particles.
Representing magnetization by single magnetic dipoles may not be sufficient any more and the magnetization within the particles should be spatially resolved to increase accuracy.

Our study aims to investigate the impact of hydrodynamic interactions beyond Stokes drag on the structure formation in magnetically driven particles.
To face the challenges outlined above, we build on a method that considers the particles as parts of the fluid, yet, with significantly enhanced viscosity that hinders their deformation~\cite{tanaka2000simulation,kodama2004fluid,tanaka2006viscoelastic,furukawa2014activity,menzel2014active,tateno2021power}. Thus, couplings between particles and fluid as well as hydrodynamic interactions are intrinsic to the method. 
Approach into virtual contact is supported. 
We extend the ``fluid particle dynamics'' (FPD) method~\cite{tanaka2000simulation,tanaka2006viscoelastic,furukawa2018physical} by the influence of externally imposed magnetic fields on magnetically soft, magnetizable particles. 
Here, the FPD framework allows for an efficient implementation when spatially resolving magnetic quantities to accurately determine the magnetic forces.
Due to its simplicity, the method can further be extended to include more complex rheologies in future studies.
This is particularly relevant for structure formation in non-Newtonian fluids.

We explore the impact of spatially resolved hydrodynamic fluid flows on structure formation by comparison with systems that reduce the effects of the surrounding carrier media to simple drag forces. 
Hydrodynamic interactions promote the formation of string-like structures already for nondirectional, isotropic interactions between mutually attracting particles~\cite{tanaka2000simulation,furukawa2010key,tanaka2006viscoelastic}. 
Thus, we expect that hydrodynamics supports the formation of string-like particulate structures also in magnetic fluids. There, 
external magnetic fields lead to anisotropic particle interactions.
Indeed, we observe the emergence of slender, string-like structures when resolving hydrodynamic flows.
Conversely, more compact aggregates result when we only work with drag forces.
Since the nature of the internal structure has significant impact on overall performance of magnetic composite systems \cite{filipcsei2007magnetic,potisk2019continuum}, this is an important aspect.

Key to the FPD method~\cite{tanaka2000simulation,tanaka2006viscoelastic,furukawa2018physical}
is to represent the presence of any particle $i$ through a continuous field $\phi_i$.
Similarly to phase-field models~\cite{steinbach2009phase,qin2010phase}, this field distinguishes between the inside and the outside of the particle, with $\phi_i = 1$ inside, $\phi_i = 0$ outside, and a smooth transition in between.
We set $\phi_i$ as
\begin{equation}
\label{eq:phi}
\phi_i(\mathbf{r}) = [\tanh\{(a-|\mathbf{r}-\mathbf{R}_i|)/c\} +1]/2 ,
\end{equation}
where $\mathbf{R}_i$ is the particle position, $a$ its radius,  and $c$ the thickness of the interface region between particle and fluid.
The particles are considered identical and as part of the fluid, albeit with significantly increased viscosity $\eta_\mathrm{p}$ compared to the fluid viscosity $\eta_\mathrm{s}$.
We define their ratio $R$ via $\eta_\mathrm{p} = R \eta_\mathrm{s}$.
Summing over all particles, we obtain the total phase field $\phi_\mathrm{tot}$ as
\begin{equation}
\label{eq:totalPhi}
\phi_\mathrm{tot}(\mathbf{r})=  \sum_{i=1}^{N} \phi_i(\mathbf{r}),
\end{equation}
where $N$ is the total number of particles.
As a function of space, the viscosity follows as
\begin{equation}
\label{eq:viscosityField}
\eta(\mathbf{r})=  \eta_\mathrm{s} + (R - 1) \eta_\mathrm{s} \phi_\mathrm{tot}(\mathbf{r}).
\end{equation}
The viscosity field enters the Navier--Stokes equation
\begin{equation}
\label{eq:NavierStokes}
\rho   \partial_t \mathbf{v} + \rho \mathbf{v} \cdot \nabla \mathbf{v} = \mathbf{f} -\nabla p + \nabla \cdot [\eta \{ (\nabla \mathbf{v}) +(\nabla \mathbf{v})^\top \}] ,
\end{equation}
which governs the dynamics of the velocity field $\mathbf{v}$.
Here, we assume that the mass density of the particles is equal to that of the fluid. Moreover, we restrict ourselves to incompressible fluids, that is, $\nabla \cdot \mathbf{v} = 0$.
The FPD method implicitly assumes no-slip boundary conditions on the particle surfaces.

The Navier--Stokes equation~(\ref{eq:NavierStokes}) further includes a force density field $\mathbf{f}$, which is determined from the forces $\mathbf{F}_i$ acting on the particles via
\begin{equation}
\label{eq:forceDensity}
\mathbf{f}(\mathbf{r}) =  \frac{1}{V_\mathrm{p}}  \sum_{i=1}^{N} \mathbf{F}_i \phi_i(\mathbf{r}) .
\end{equation}
$V_\mathrm{p}$ is the particle volume in three dimensions or area in two dimensions.
The force acting on each particle, thus transmitted to the fluid, is equally distributed over the nondeformable particle body.
We distinguish between forces stemming from a pair potential, characterizing, for example, steric effects, and magnetic forces between particles.
For the former, we here use the conventional Lennard--Jones potential \cite{tanaka2000simulation}, which as a function of the center-to-center distance $r$ between two particles reads
\begin{equation}
\label{eq:LennardJones}
V(r) = \epsilon \left[  \left(\frac{\sigma}{r}\right)^{12} - \left(\frac{\sigma}{r}\right)^{6} \right].    
\end{equation}
$\epsilon$ is the interaction strength and $\sigma = 2a$ determines the interaction range.

\begin{figure*}
\includegraphics[width=0.999\linewidth]{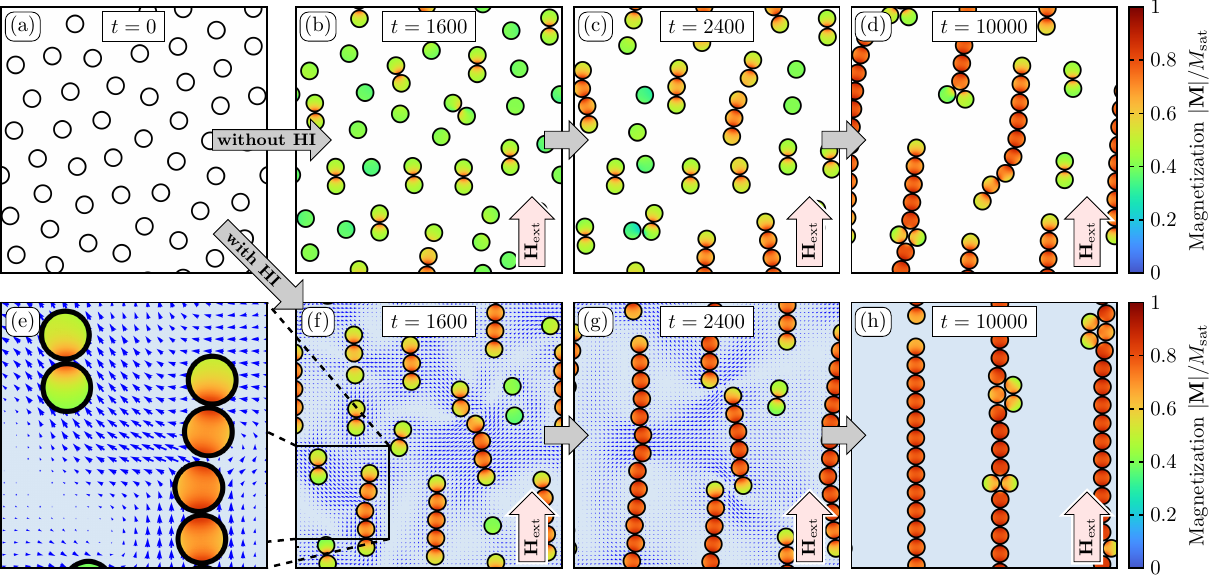}
\caption{\label{fig:snapshotsLJ_050}Structure formation with and without hydrodynamic interactions (HI) for an area fraction of $\theta \approx 0.15$. (a) Simulations start from the same configuration of roughly equally distributed particles. (b--d) Snapshots of the ``dry''  system at times $t=1600$, $2400$, and $10000$. The color scale refers to the magnetization field within the particles. (f--h) Snapshots of the system involving hydrodynamic interactions and fluid flows at the same times. Blue arrows indicate the flow field. (e) Magnified  part of (f) showing the emergence of large-scale flows and the mutual amplification of local magnetization for particles that are close to each other. The external field strength is set to $|\mathbf{H}_\mathrm{ext}|/M_\mathrm{sat} = 0.2$. (Multimedia available online)}
\end{figure*}

We extend the method to include magnetic forces between the particles.
Often, these are described via dipole--dipole interactions based on the far-field contributions to the induced magnetic fields~\cite{jaeger2013dynamics,romeis2016elongated,weeber2018polymer,dobroserdova2022magneto}.
However, during structure formation, particles approach each other until basically in contact.
Thus, more refined methods are required, which accurately describe magnetic effects even for small distances between particles.
Therefore we spatially resolve magnetization within the particles.
Utilizing the phase fields $\phi_i$,
the magnetization field $\mathbf{M}$ is determined from the local magnetic field $\mathbf{H}$ using the Langevin function
\begin{equation}
\label{eq:magnetization}
\mathbf{M} =  M_\mathrm{sat}     \left[  \coth(\alpha \phi_\mathrm{tot} |\mathbf{H}|)   - \frac{1}{\alpha \phi_\mathrm{tot} |\mathbf{H}|}    \right] \frac{\mathbf{H}}{|\mathbf{H}|}.
\end{equation}
Here, $M_\mathrm{sat}$ denotes the magnitude of the saturation magnetization.
The linearized relation between $\mathbf{M}$ and $\mathbf{H}$ for weak fields is determined by $\mathbf{M} = \alpha M_\mathrm{sat} \mathbf{H}/3$~\cite{puljiz2018reversible}.
Thus, the parameter $\alpha$ is directly connected to the susceptibility.
Such Langevin-type models are appropriate for magnetically soft particles.
We here focus on materials with a very fast relaxation time of the magnetization process compared to the time scale of their induced translational motion and overall structure formation.
The magnetization within the particles is thus assumed to be instantaneously determined by the magnetic field. 
We use Eq.~(\ref{eq:magnetization}) to locally evaluate the resulting magnetization and magnetic fields within the particles.
The quantities $\alpha$ and $M_\mathrm{sat}$ are fitted to experimental curves, for example, for nickel~\cite{puljiz2018reversible} or carbonyl iron~\cite{kalina2020macroscopic} particles, which are commonly used~\cite{vicente2011magnetorheological}.
Previous studies~\cite{puljiz2018reversible,kalina2020macroscopic} have demonstrated that the approach accurately describes the resulting mechanical behavior of magnetically soft, micron-sized particles.

The magnetic field $\mathbf{H}$, in turn, is composed of an applied uniform external magnetic field $\mathbf{H}_\mathrm{ext}$ and the field induced by the magnetized particles themselves.
To determine this additional contribution, we solve Maxwell's equations in space, that is,
\begin{equation}
\label{eq:Maxwell}
\bm{\nabla} \cdot \mathbf{B} = 0, \qquad 
\bm{\nabla} \times \mathbf{H} = \mathbf{0},
\end{equation}
where $\mathbf{B}$ and $\mathbf{H}$ are related via
\begin{equation}
\mathbf{B} = \mu_0 (\mathbf{M}+\mathbf{H}).
\end{equation}
Having obtained $\mathbf{M}$ and $\mathbf{B}$, the magnetic force density can be calculated via $\mathbf{f}^\mathrm{mag} = \bm{\nabla} \mathbf{B}\cdot \mathbf{M} $~\cite{degroot1972foundations,eringen2012electrodynamics,metsch2016numerical,metsch2019two}.
The magnetic forces acting on the particles are obtained as 
\begin{equation}
\label{eq:magneticForces}
\mathbf{F}^\mathrm{mag}_i = \int \mathrm{d}\mathbf{r}\  \bm{\nabla}  \mathbf{B}\cdot\mathbf{M}_i  .
\end{equation}
$\mathbf{M}_i$ is the magnetization field within particle $i$, calculated in analogy to Eq.~(\ref{eq:magnetization}), but using the phase field $\phi_i$ instead of $\phi_\mathrm{tot}$.

To determine the actual motion of the particles, we return to the discrete particle level.
The particle velocities $\mathbf{V}_i$ are obtained as averages of the velocity field $\mathbf{v}$ over the particle bodies, 
\begin{equation}
\label{eq:particleVelocities}
\mathbf{V}_i = \frac{1}{V_\mathrm{p}} \int \mathrm{d}\mathbf{r}\   \mathbf{v}(\mathbf{r}) \phi_i(\mathbf{r}).
\end{equation}
Time integration of $\mathbf{V}_i$ updates the particle positions. 

Here, we focus on microscopic particles of a few or tens of micrometers (or more) in size, which are used in magnetorheological fluids~\cite{vicente2011magnetorheological}.
Thus, we neglect thermal fluctuations. Extending the simulation method to particles on the nanometer scale would necessitate taking into account thermal effects.

In practice, the different steps of the FPD method are performed in sequence. 
We first calculate the forces acting on the particles, including the magnetic forces 
as described above.
These enter the calculations of the flow field, which is performed using an implicit Euler scheme combined with a pseudospectral method~\cite{canuto2007spectral}.
This efficient approach evaluates spatial gradients in Fourier space and makes use of the fast Fourier transform to speed up the calculations.
Particle positions are then updated according to Eq.~(\ref{eq:particleVelocities}). Afterwards, the sequence repeats.
We refer to the Supplementary Material for further details.

To evaluate the equations numerically, we rescale our quantities.
We use the particle radius $a$ as a length scale and introduce the time scale $\sqrt{\rho a^5/\epsilon}$, which involves the strength of the Lennard--Jones potential.
In combination, the unit of velocity becomes $\sqrt{\epsilon/(\rho a^3)}$.
The strength of viscous forces is determined by $C_\mathrm{visc} = \eta_\mathrm{s} \sqrt{a /(\rho\epsilon)}$, which we set $C_\mathrm{visc} = 1$. 
This leads to observed local Reynolds numbers $\mathrm{Re} \approx 10^{-2}$.
Thus, inertial effects are not expected to play a significant role.
Magnetization $\mathbf{M}$ and magnetic field $\mathbf{H}$ are both rescaled by $M_\mathrm{sat}$, while the magnetic flux density $\mathbf{B}$ is measured in units of $\mu_0 M_\mathrm{sat}$.
The strength of magnetic forces is then characterized by  $C_\mathrm{mag} = \mu_0 M_\mathrm{sat}^2 a^3/\epsilon$.
We set $C_\mathrm{mag} = 1$ so that the strength of Lennard--Jones and magnetic forces are comparable.
The magnetic parameters are set to $M_\mathrm{sat} = \SI{3.333e5}{\ampere\per\meter}$ and $\alpha = \SI{1.179e-4}{\meter\per\ampere}$, corresponding to paramagnetic nickel particles~\cite{puljiz2018reversible}.

Concerning the numerical method, the values chosen for the viscosity ratio $R$ and the interface thickness $c$ represent a tradeoff between the accuracy of the method and numerical efficiency.
In particular, $R = 50$ has been shown to provide a reasonable balance~\cite{tanaka2000simulation}.
We further use an interface thickness of $c = 0.2a$ for the hydrodynamic calculations and a smaller value of $c=0.1a$ for the magnetic calculations to be able to resolve the magnetization within the particles accurately.

More details on rescaling and numerical parameters are found in the Supplementary Material.
In the following, we set the strength of the external magnetic field to $|\mathbf{H}_\mathrm{ext}| = 0.2 M_\mathrm{sat}$.
Average magnetization thus remains below the saturation value.

An extensive exploration of the parameter space is beyond the scope of this study. 
We rather focus on a specific set of parameters and particle material with the aim of  identifying and illustrating the impact of hydrodynamic effects beyond Stokes drag on the structure formation in planar systems.
To this end, we compare our simulations that include fluid flows and thus mutual hydrodynamic interactions between the particles via Eq.~(\ref{eq:NavierStokes}) to discrete particle dynamics in the absence of explicit fluid interactions.
In the latter ``dry'' case, particle velocities $\mathbf{V}_i$ are not determined from the flow field, but directly from Newton's second law with an effective linear friction parameter $\gamma$,
\begin{equation}
\label{eq:underdampedDynamics}
\frac{\mathrm{d}\mathbf{V}_i}{\mathrm{d}t} = \mathbf{F}^\mathrm{mag}_i + \mathbf{F}^\mathrm{LJ}_i - \gamma \mathbf{V}_i,
\end{equation}
in rescaled units. 
$\mathbf{F}^\mathrm{LJ}_i$ are the forces due to the Lennard--Jones potential. 
We set $\gamma = 50$, which results in an aggregation time scale comparable to structure formation under hydrodynamic couplings.
The common assumption of reducing the effects of the carrier medium to effective Stokes drag on the particles results in qualitatively the same friction term as in Eq.~(\ref{eq:underdampedDynamics}).

\begin{figure*}
\includegraphics[width=0.999\linewidth]{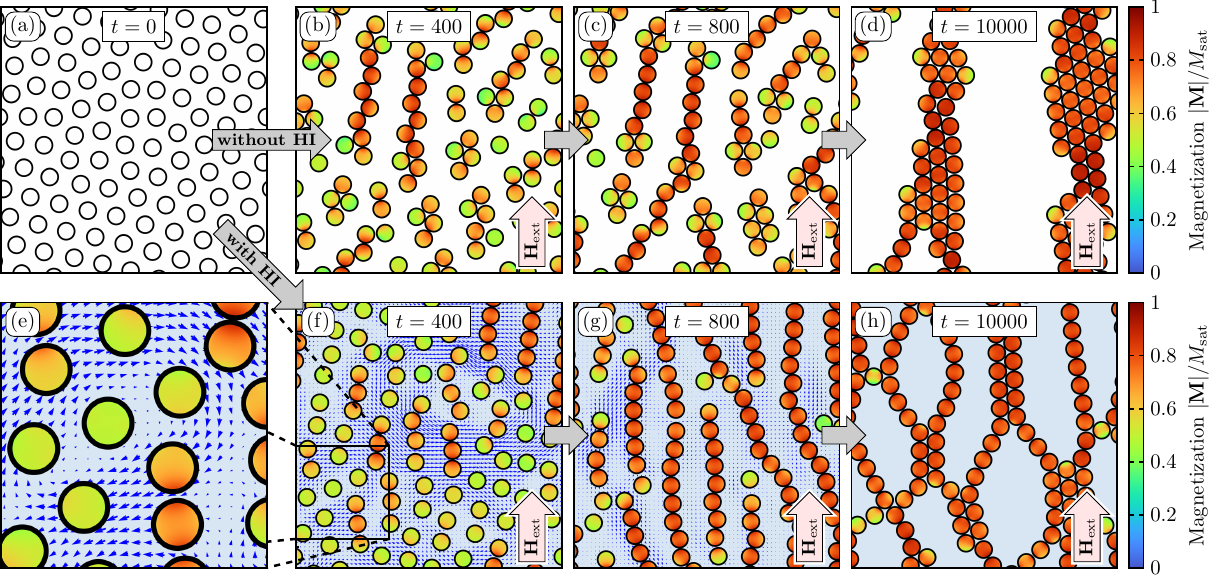}
\caption{\label{fig:snapshotsLJ_100}
Same as in Fig.~\ref{fig:snapshotsLJ_050}, yet for an elevated area fraction of $\theta \approx 0.31$ and snapshots in (b--d) and (f--h) taken at times $t=400$, $800$, and $10000$. Here, chain formation promoted by hydrodynamic interactions (HI) leads to branching and string-like structures when compared to the more compact clusters that emerge without hydrodynamics. (Multimedia available online)}
\end{figure*}

Focusing on planar systems and evaluating the equations in two dimensions, our numerical simulations represent monolayers of particles.
Here, we are interested in two-dimensional bulk behavior in the absence of lateral boundaries and, thus, apply periodic boundary conditions.
We start from initial configurations of roughly equally spaced particles, see Fig.~\ref{fig:snapshotsLJ_050}(a) for a snapshot (Multimedia available online).
Once the magnetic field is turned on, the particles become magnetized.
Magnetization, in turn, results in magnetic forces between particles, see Eq.~(\ref{eq:magneticForces}).
Figure~\ref{fig:snapshotsLJ_050} shows snapshots for a low area fraction of $\theta \approx 0.15$ at different times, see Supplementary Movie 1 for a dynamic visualization.
Both the magnetization within the particles as well as the flow field in the hydrodynamic case are visualized.
Chains are formed in either case.
However, chain formation is promoted by hydrodynamic coupling.
Large-scale flows emerge that facilitate movement.
For example, in Fig.~\ref{fig:snapshotsLJ_050}(g) a chain of six particles is transported as one unit to connect with another chain into a larger structure.
Without explicit hydrodynamics, formation of larger chain-like structures is less pronounced, see the upper row of Fig.~\ref{fig:snapshotsLJ_050}.
The snapshots also show the strong mutual reinforcement of local magnetization for particles close to each other and oriented parallel to the external magnetic field, see Fig.~\ref{fig:snapshotsLJ_050}(e).
This collective magnetic effect may additionally accelerate the aggregation of chains, specifically below magnetization saturation.

Similar effects can be observed for larger area fractions of particles.
Figure~\ref{fig:snapshotsLJ_100} shows snapshots for an area fraction of $\theta \approx 0.31$, again comparing the ``dry'' case with the situation under hydrodynamic coupling, see also Supplementary Movie 2 (Multimedia available online).
In the beginning, the absence of hydrodynamics leads to faster aggregation, compare Fig.~\ref{fig:snapshotsLJ_100}(b) and (f). However, chain formation is accelerated under hydrodynamic interactions once large-scale flows have formed.
In Fig.~\ref{fig:snapshotsLJ_100}(f), vortex flows occupy larger parts of the system, as magnified in Fig.~\ref{fig:snapshotsLJ_100}(e).

Figure~\ref{fig:snapshotsLJ_100}(d) and (h) show the developed structures after a longer time.
Comparing the two cases, we observe a crucial difference.
With hydrodynamic couplings, branching, string-like structures have formed. Chains of particles are on average aligned along the external magnetic field.
In the absence of hydrodynamic effects, the emerging aggregates are much more compact. They consist of multiple layers of hexagonally arranged particles.
This is in line with previous observations of particle aggregation under attractive isotropic interactions~\cite{tanaka2000simulation,furukawa2010key,tanaka2006viscoelastic}.

\begin{figure}
\includegraphics[width=0.999\linewidth]{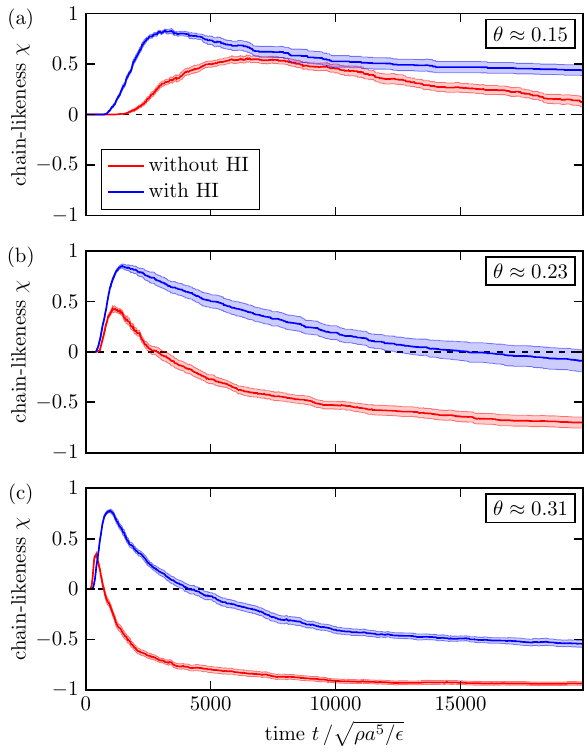}
\caption{\label{fig:chainLikeness}Evolution of the chain-likeness parameter $\chi$ for an area fraction of (a) $\theta \approx 0.15$, (b) $\theta \approx 0.23$, and (c) $\theta \approx 0.31$ and an external field strength of $|\mathbf{H}_\mathrm{ext}| = 0.2 M_\mathrm{sat}$. Hydrodynamic interactions (HI) consistently lead to more chain-like structures when compared to the ``dry'' case without HI. Shaded areas indicate standard errors, that is, standard deviations divided by the square root of the sample size.}
\end{figure}

To further examine the structure formation, we introduce a chain-likeness parameter $\chi$.
It quantifies the relative number of particles in a two-neighbor configuration, that is, in slender chains of one particle in diameter, compared to more compact clusters where particles have a higher number of neighbors.
We define $\chi$ via
\begin{equation}
\label{eq:chainLikeness}
\chi = \frac{N_2 - (N_3 + N_4 + N_5 + N_6)}{N}.
\end{equation}
$N_j$ is the number of particles with $j$ neighbors in the system. 
Particles are counted as neighbors when the distance between their centers is smaller than a threshold value, set to $1.2\sigma$. This corresponds to a gap between particles of below $\SI{20}{\percent}$ of the particle diameter. 
We varied the threshold value and did not observe any qualitative differences in the results for values between $1.2\sigma$ and $1.5\sigma.$
A configuration of perfect, nonbranching chains leads to $\chi = 1$, whereas completely separated particles lead to $\chi = 0$.
Thus, for example, the observed string-like structures in Fig.~\ref{fig:snapshotsLJ_100}(g) would yield a relatively high value.
Negative values indicate that more compact clusters have formed, where particles have on average more than two neighbors.
Such structures include, for example, the multi-layer, hexagonally ordered aggregates in Fig.\ref{fig:snapshotsLJ_100}(d).
Figure~\ref{fig:chainLikeness} shows the evolution of $\chi$ for three different area fractions of particles. 
In each case, we repeated the calculations starting from $12$ different initial configurations and averaged over the results.

We first consider the small area fraction $\theta \approx 0.15$.
Although slower in the absence of hydrodynamic couplings, both cases, ``wet'' and ``dry'', exhibit the formation of chain-like aggregates in the beginning, indicated by increasing $\chi$.
After some time, $\chi$ decreases as a result of some of the chains combining into multiple-layer structures.
Throughout, both the maximum and the average chain-likeness is significantly higher in the ``wet'' case under hydrodynamic coupling.

Increasing the area fraction to $\theta \approx 0.23$, we still observe an initial stage of chain formation.
However, the subsequent decrease of $\chi$ is both faster and stronger.
In the ``dry'' case, $\chi$ quickly dips below zero, indicating the predominance of more compact clusters.
Conversely, hydrodynamic couplings lead to much slower decrease of $\chi$. Yet, finally, negative values likewise result. 

Elevated area fractions of $\theta \approx 0.31$ further amplify these effects.
$\chi$ decreases very quickly to negative values when hydrodynamic interactions are absent. 
It approaches $\chi = -1$, reflecting the formation of larger clusters, see Fig.~\ref{fig:snapshotsLJ_100}(d).
Turning to the case with hydrodynamics included, we observe a similar process, although much slower.
The value of $\chi$ at larger times is consistently higher, in line with branched, but still string-like structures as in Fig.~\ref{fig:snapshotsLJ_100}(h).

Overall, 
the maximum of $\chi$ is consistently larger when hydrodynamic effects are included, due to the formation of string-like chains in contrast to more compact clustering.
The process of chain formation seems to be more consistent as well.
In particular, the maximum of $\chi$ is always about the same height and reached after more comparable times. 
Conversely, the area fraction has a significantly higher impact on the dynamics in the ``dry'' system. 
Here, the initial chain formation is much faster at high area fraction compared to the slow aggregation at low area fraction.
This more pronounced difference is in line with the absence of large-scale flows, which promote particle aggregation even for larger initial distances between particles under hydrodynamic coupling.

As observed before~\cite{tanaka2006viscoelastic},  particles tend to arrange into string-like branching structures when explicit hydrodynamic interactions are considered.
This is due to hydrodynamic many-particle effects beyond single-particle Stokes drag.
For illustration, we mention two such phenomena.
First, when two particles approach each other, the fluid in between is squeezed outward in the perpendicular direction.
This flow transports particles in the perpendicular vicinity away~\cite{tanaka2006viscoelastic}.
Second, the mobility of a pair of particles or of a larger chain is much lower in the direction perpendicular to the chain axis than along it~\cite{tanaka2006viscoelastic}.
These and similar effects significantly influence the assembly process of aggregates and tend to favor the string-like structures we observe.

Since hydrodynamic interactions only play a role during the motion of the particles and vanish otherwise, the structures in the long-time limit are expected to be qualitatively similar with and without explicit hydrodynamics, if in equilibrium.
As the snapshots in Fig.~\ref{fig:snapshotsLJ_100} and the evolution of $\chi$ in Fig.~\ref{fig:chainLikeness} indicate, compact, multi-layer aggregates tend to form in the long-time limit for both cases.
However, we have demonstrated that the assembly pathway is strongly affected by the presence of hydrodynamic interactions. In particular, string-like branching structures emerge as long-lasting intermediate states.

\begin{figure}
\includegraphics[width=0.999\linewidth]{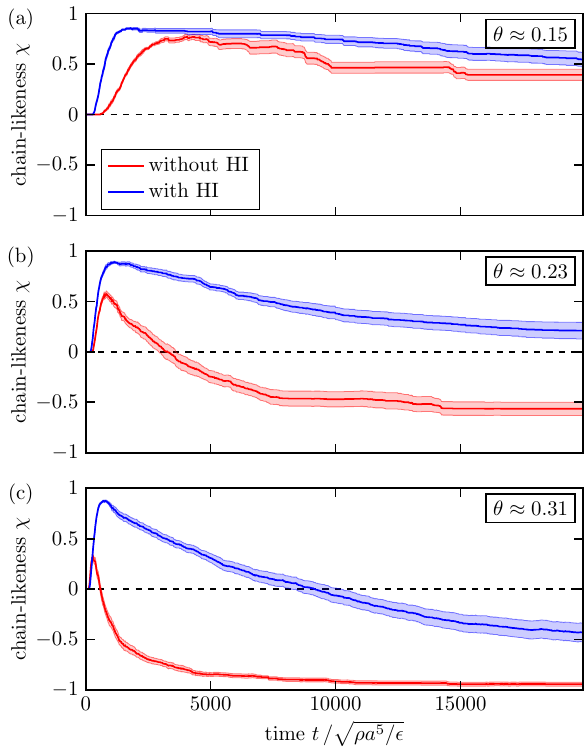}
\caption{\label{fig:chainLikenessSaturated}Evolution of the chain-likeness parameter $\chi$, in analogy to Fig.~\ref{fig:chainLikeness}, yet for an increased magnitude of the external magnetic field $|\mathbf{H}_\mathrm{ext}| = M_\mathrm{sat}$. Here, saturation effects become relevant and the importance of higher-order effects beyond dipole-dipole interactions is reduced. We observe the same qualitative behavior as for the smaller $|\mathbf{H}_\mathrm{ext}|$ in Fig.~\ref{fig:chainLikeness}.}
\end{figure}

Concerning the impact of the magnetization behavior on the types of observed structure formation, we remark that we observed qualitatively similar behavior for other regimes of the magnetization curve as well.
For instance, we increased the external magnetic field to $|\mathbf{H}_\mathrm{ext}| = M_\mathrm{sat}$.
Here, saturation effects of particle magnetization become generally relevant.
As a result, the importance of higher-order effects beyond dipole-dipole interactions is reduced for our spherical particles.
Repeating the simulations, we find the same qualitative behavior.
This is illustrated by Fig.~\ref{fig:chainLikenessSaturated}, which shows the evolution of the chain-likeness parameter.
In particular, emerging hydrodynamic flows favor the emergence of string-like aggregates compared to the more compact clusters in the absence of explicit hydrodynamic effects.

In view of the general importance of magnetic interactions beyond the dipole approximation, as discussed by previous studies~\cite{romeis2026beyond,biller2014modeling}, they may significantly affect the magnetic interactions, specifically below magnetization saturation.
The simulation method we employ spatially resolves the magnetization within the particles and, thus, captures these effects.
This is of particular importance for prospective extensions of the method, for example, considering anisotropic particle shapes.

Summarizing, we extended the fluid particle dynamics method~\cite{tanaka2000simulation,tanaka2006viscoelastic} to investigate magnetically induced structure formation in magnetorheological fluids.
Spatially resolved and mutual particle magnetization together with resulting elevated magnetic forces between particles are included. These effects are important below saturation magnetization.
Specifically, we compare magnetically driven structure formation with and without hydrodynamic couplings. We find qualitative differences. Hydrodynamics promotes the formation of slender oriented chains and string-like structures, opposed to the more compact clusters emerging in its absence. In this way, hydrodynamic interactions serve to enhance structure formation in magnetorheological composite systems. 

Besides providing a fundamental understanding of the background of structure formation in magnetic fluids in general, our results have practical implications when fabricating structured magnetorheological elastomers of improved properties. They are generated by crosslinking suspensions of magnetic particles in the presence of external magnetic fields to lock in chain-like particle aggregates~\cite{carlson2000mr,vicente2011magnetorheological,gunther2012x}.
For example, the maximum in the chain-likeness parameter during structure formation 
implies an optimal time for crosslinking that leads to the most pronounced slender, chain-like structures.
It is mandatory to understand the process of structure formation. Then, we can influence it in a requested way, which has direct impact on the overall magnetic, electric, and mechanical properties \cite{filipcsei2007magnetic,mietta2016anisotropic}.

The numerical simulations are performed for two-dimensional systems, representing monolayers of particles.
To experimentally validate our results, we suggest experiments that isolate the impact of hydrodynamic interactions in a similar way.
In our case, we have neglected frictional effects of the flow with a substrate, although it can be qualitatively mimicked in a straightforward way, for instance, by an additional term of linear friction~\cite{brinkman1949permeability,vilquin2021asymptotic,daddi2024hydrodynamics,daddi2025force}. Thus, for quantitative comparison, monolayers of particles sedimented to the bottom of a freely hanging droplet provide a first approach~\cite{ebert2009experimental}. Yet, to introduce two-dimensional fluid flows, the droplet would need to be reduced to a free-standing film. Qualitatively, however, to demonstrate the basic effect of hydrodynamic interactions, we indeed suggest to compare results for a simple experimental set-up of low friction of the particles with the substrate to one of strong friction-like interactions. As one of our subsequent steps, we consider the extension of our investigations to three-dimensional systems. Then, explicit quantitative comparison with three-dimensional experimental systems is rather immediate.

\section*{Supplementary Material}

The Supplementary Material discusses how the equations introduced in the main text are rescaled for the purposes of numerical calculations and gives more details on the numerical methods we employ. It also includes additional information on the Supplementary Movies.

\begin{acknowledgments}
The authors thank Karl Kalina for stimulating discussions.  
For funding, the authors thank the Deutsche Forschungsgemeinschaft (German Research Foundation, DFG) for support through Research Unit FOR 5599 on structured magnetic elastomers, project nos.\ 511114185, 535421963, and 535543971. 
Further support (T.A.) was provided by JSPS KAKENHI Grant Number 25K00967.
\end{acknowledgments}

\section*{Author Declarations}
\subsection*{Conflict of Interest}
The authors have no conflicts to disclose.

\subsection*{Author Contributions}

\textbf{Henning Reinken}: Conceptualization (supporting); Data curation (equal); Formal analysis (lead);
Investigation (lead); Methodology (equal); Software (equal); Supervision (supporting);
Validation (equal); Visualization (lead); Writing – original draft (lead); Writing – review and editing (equal).

\textbf{Markus Heiber}: Data curation (equal); Formal analysis (supporting); Investigation (supporting); Methodology (equal); Software (equal); Visualization (supporting); Writing – original draft (supporting).

\textbf{Takeaki Araki}: Methodology (supporting); Validation (equal); Writing – review and editing (equal).

\textbf{Andreas M. Menzel}: Conceptualization (lead); Funding acquisition; Methodology (equal); Project administration; Resources; Supervision (lead); Validation (equal); Writing – review and editing (equal).

\section*{Data Availability Statement}

The data that support the findings of this study are available from the corresponding author upon reasonable request.

\bibliography{references}

\end{document}



\title{Hydrodynamics substantially affects induced structure formation in magnetic fluids\\[0.5\baselineskip]
\textit{Supplementary Material\vspace*{0.5cm}}}

\author{Henning Reinken}
\email{henning.reinken@ovgu.de}
\affiliation{Otto-von-Guericke-Universität Magdeburg, Institut f\"ur Physik, Universitätsplatz 2, 39106 Magdeburg, Germany} 

\author{Markus Heiber}
\affiliation{Otto-von-Guericke-Universität Magdeburg, Institut f\"ur Physik, Universitätsplatz 2, 39106 Magdeburg, Germany}  

\author{Takeaki Araki}
\affiliation{Kyoto University, Department of Physics, Kyoto 606-8502, Japan} 

\author{Andreas M. Menzel}
\email{a.menzel@ovgu.de}
\affiliation{Otto-von-Guericke-Universität Magdeburg, Institut f\"ur Physik, Universitätsplatz 2, 39106 Magdeburg, Germany}

\date{\today}

\begin{abstract}
This Supplementary Material discusses how the equations introduced in the main text are rescaled for the purposes of numerical calculations, gives more details on the numerical methods we employ and includes additional information on the Supplementary Movies.
\end{abstract}

\pacs{}

\maketitle 

\section{Rescaling}

We rescale the equations to reduce the number of free parameters.
To this end, we employ the particle radius $a$ as spatial scale and utilize the strength $\epsilon$ of the Lennard--Jones potential to define a time scale. Thus, we rescale space and time via
\begin{equation}
x = a \tilde{x}, \qquad t = \sqrt{\frac{a^5 \rho}{\epsilon}} \, \tilde{t},
\end{equation}
where the tilde indicates rescaled units.
The rescaled Navier--Stokes equation is then obtained as
\begin{equation}
\label{eq:NavierStokesRescaled}
\partial_{\tilde{t}} \tilde{\mathbf{v}} + \tilde{\mathbf{v}} \cdot \tilde{\bm{\nabla}} \tilde{\mathbf{v}} = \tilde{\mathbf{f}} - \tilde{\bm {\nabla}} \tilde{p} + C_\mathrm{visc} \tilde{\bm{\nabla}} \cdot [\tilde{\eta} \{ (\tilde{\bm{\nabla}} \tilde{\mathbf{v}}) +(\tilde{\bm{\nabla}} \tilde{\mathbf{v}})^\top \}] ,
\end{equation}
with $\tilde{\eta}(\tilde{\mathbf{r}}) = 1 + \sum_{i=1}^{N}  (R-1)
\phi_i(\tilde{\mathbf{r}})$ and $\tilde{p} = a^3 p / \epsilon$.
The strength of viscous forces compared to the inter-particle potential is characterized by 
$C_\mathrm{visc} = \eta_\mathrm{s} \sqrt{a/(\rho\epsilon)}$.
Throughout the work, we set $C_\mathrm{visc} = 1$.
We determine particle-based Reynolds numbers $\mathrm{Re}_i$, employing the particle diameter $2a$, the viscosity $\eta_\mathrm{s}$ of the surrounding fluid and the current particle velocity $\mathbf{V}_i$ via $\mathrm{Re}_i = 2 a \rho |\mathbf{V}_i| / \eta_\mathrm{s}$.
Maximum observed Reynolds numbers are of the order of $\mathrm{Re}_i \approx 10^{-2}$.
Thus, for the chosen value of $C_\mathrm{visc}$, viscous effects dominate over inertia.

The magnetization $\mathbf{M}$, magnetic flux density $\mathbf{B}$, and magnetic field $\mathbf{H}$ are rescaled using the saturation magnetization $M_\mathrm{sat}$ and vacuum magnetic permeability $\mu_0$, 
\begin{equation}
\mathbf{M} = M_\mathrm{sat} \tilde{\mathbf{M}}, \quad \mathbf{H} = M_\mathrm{sat} \tilde{\mathbf{H}}, \quad \mathbf{B} = \mu_0 M_\mathrm{sat} \tilde{\mathbf{B}}.
\end{equation}
Consequently, the rescaled equation for the magnetization reads
\begin{equation}
\label{eq:magnetization}
\tilde{\mathbf{M}} = \left[\coth(\tilde{\alpha} \phi_\mathrm{tot} |\tilde{\mathbf{H}}|)   - \frac{1}{\tilde{\alpha} \phi_\mathrm{tot}  |\tilde{\mathbf{H}}|}    \right] \frac{\tilde{\mathbf{H}}}{|\tilde{\mathbf{H}}|},
\end{equation}
where $\tilde{\alpha} = \alpha M_\mathrm{sat}$.
Throughout our work, we use $\tilde{\alpha} = 39.3$. 
This approximately corresponds to $M_\mathrm{sat} = \SI{3.333e5}{\ampere\per\meter}$ and $\alpha = \SI{1.179e-4}{\meter\per\ampere}$, which is appropriate for nickel particles~\cite{puljiz2018reversible}.
Magnetic forces entering the force density $\tilde{\mathbf{f}}$ in the Navier--Stokes equation are calculated in rescaled units via 
\begin{equation}
\label{eq:magneticForcesRescaled}
\tilde{\mathbf{F}}_i^\mathrm{mag} =C_\mathrm{mag} \int \mathrm{d}\tilde{\mathbf{r}}\  \tilde{\bm{\nabla}}  \tilde{\mathbf{B}}\cdot \tilde{\mathbf{M}}_i .
\end{equation}
Their overall strength compared to the inter-particle potential is characterized by 
$C_\mathrm{mag} = \mu_0 M_\mathrm{sat}^2 a^3/\epsilon$.
We set $C_\mathrm{mag} = 1$ throughout.

\section{Numerical methods}

The simulation method of fluid particle dynamics~\cite{tanaka2000simulation} involves a combination of a discrete particle picture and a hydrodynamic field description.
Connections between these two levels of description are based on phase-fields $\phi_i$ to represent the presence of each particle $i$. They vary smoothly from the inside to the outside of the particles.

All fields are evaluated numerically, using grids with different resolution for the hydrodynamic quantities such as the velocity field $\mathbf{v}$ and the magnetic quantities such as the magnetization $\mathbf{M}$ and the magnetic field $\mathbf{H}$.
This approach allows for a very efficient implementation.
To obtain the results discussed in this work, we use a spatial resolution of $320\times 320$ grid points for the hydrodynamic calculations.
The interface thickness of the transition region, see Eq.~(1) in the main text, is set to $c = 0.2a$ for the hydrodynamic calculations.
This value represents a tradeoff between accuracy and numerical efficiency as the interface has to be fully resolved by the spatial discretization.
For the magnetic quantities, we use a resolution of $640\times 640$ grid points and $c = 0.1a$.
This small value allows us to resolve the magnetization accurately even close to the boundaries of the particles.
The factor $R$, which characterizes how strongly the viscosity is increased within the particles, see Eq.~(3) in the main text, is set to $R = 50$.
This value was shown to be sufficient in previous studies~\cite{tanaka2000simulation}.

We utilize a pseudospectral method~\cite{canuto2007spectral}, that is, spatial derivatives are calculated in Fourier space, which benefits from the efficient implementation of the Fast Fourier Transform.
Throughout this work, we consider periodic boundary conditions.
We fix the system size to $32a \times 32a$ and vary the number of particles.
Specifically, we choose $50$, $75$, and $100$ particles, which yields area fractions of $\theta \approx 0.15$, $\theta \approx 0.23$, and $\theta \approx 0.31$.

As summarized in the main text, the basic procedure in every time step involves, first, the calculation of magnetic and other forces between particles. Second, these enter the hydrodynamic description as a force density.
Third, the resulting velocity field then leads to particle motion, producing an updated particle configuration. 
In the following, we outline these steps in more detail.

In every time step, we obtain the magnetization field $\mathbf{M}$, magnetic flux density $\mathbf{B}$, and magnetic field $\mathbf{H}$ by simultaneously solving Eq.~(\ref{eq:magnetization}) and the Maxwell equations, see Eqs.~(8) in the main text.
Here, we distinguish between the uniform external magnetic field $\mathbf{\mathbf{H}_\mathrm{ext}}$ and the additional inhomogeneous internal magnetic field $\mathbf{H}_\mathrm{inh}$ induced by the magnetization of the particles themselves. The total magnetic field is given as the sum of both terms, $\mathbf{H} = \mathbf{\mathbf{H}_\mathrm{ext}} + \mathbf{H}_\mathrm{inh}$.
To solve Maxwell's equations, we further introduce a scalar potential $\psi$ for the inhomogeneous part of the magnetic field, $\mathbf{H}_\mathrm{inh} = - \bm{\nabla} \psi$.
As a result, $\mathbf{H}$ always satisfies Amp\`ere's law, $\bm{\nabla} \times \mathbf{H} = \mathbf{0}$. 
Substituting $\mathbf{B} = \mu_0(\mathbf{M} + \mathbf{H})$ into Gauss's law, $\bm{\nabla} \cdot \mathbf{B}=0$, we obtain
\begin{equation}
\label{eq:scalarPotentialFromM}
\nabla^2 \psi = \bm{\nabla} \cdot \mathbf{M},
\end{equation}
which is solved in Fourier space.

These equations are combined in an iterative procedure that involves the following steps.
We first obtain an initial iteration of the magnetization field via Eq.~(\ref{eq:magnetization}) based on the magnetic field $\mathbf{H}$ of the last step.
Then, we calculate the corresponding scalar potential $\psi$ via Eq.~(\ref{eq:scalarPotentialFromM}) from the magnetization field $\mathbf{M}$.
It provides an updated magnetic field, which we then use to correct the magnetization field, starting the next step of iteration.
To guarantee stability, we further introduce an under-relaxation procedure in the calculation of $\mathbf{M}$.
As a termination condition, we track the relative local change in the magnetization field and consider the solution to be converged once the maximum local change per iteration step is below $\SI{0.1}{\percent}$. 
For the calculation of the magnetic forces via Eq.~(\ref{eq:magneticForcesRescaled}), we need the magnetic flux density, which is readily obtained from the magnetization $\mathbf{M}$ and magnetic field $\mathbf{H}$ via $\mathbf{B} = \mu_0(\mathbf{M} + \mathbf{H})$.

All magnetic forces between particles as well as the Lennard--Jones forces are combined into a single force density for the current particle configuration via Eq.~(5) in the main text. It then enters the hydrodynamic calculations.
The Navier--Stokes equation is discretized in time via the implicit Euler method.
For every time step, this yields an implicit equation for the velocity field in the new time step, which we solve iteratively in Fourier space via an under-relaxation procedure.
Analogously to the iteration procedure in the context of the magnetic quantities, we track the relative local change in the velocity field and consider the solution to be converged once the maximum local change per iteration step is below $\SI{0.1}{\percent}$. 
Simultaneously, the pressure is determined such that the velocity field stays divergence-free and the incompressibility condition is fulfilled~\cite{canuto2007spectral}.


Finally, having obtained the velocity field, we go back to the discrete particle level by calculating the discrete particle velocities $\mathbf{V}_i$ for every particle $i$ via Eq.~(11) in the main text.
Then, the particle positions are updated by an explicit Euler step as
\begin{equation}
\mathbf{R}_i(t + \Delta t) = \mathbf{R}_i(t) + \Delta t \mathbf{V}_i(t).
\end{equation}
Using the new particle configuration, the fields $\phi_i$ are then updated as well, which completes the time step.

In the case without explicit hydrodynamic interactions, the many-particle system is governed by underdamped dynamics, including a linear friction term, see Eq.~(12) in the main text.
Here, we calculate the particle velocities and updated positions directly from the current forces via an explicit Euler procedure, 
\begin{equation}
\begin{aligned}
\mathbf{V}_i(t + \Delta t) &= \mathbf{V}_i(t) + \Delta t \big[\mathbf{F}_i(t) - \gamma \mathbf{V}_i(t)\big],\\
\mathbf{R}_i(t + \Delta t) &= \mathbf{R}_i(t) + \Delta t \mathbf{V}_i(t).
\end{aligned}
\end{equation}

For the case with full hydrodynamic interactions, the time step is set to $\Delta t = 2$, whereas for the dynamics without hydrodynamic interactions, we have to use a much smaller step $\Delta t = 0.01$ due to the explicit Euler method. 
To speed up the calculations, the magnetization field and magnetic forces are only recalculated every $100$ time steps in the case without hydrodynamics, which is still double the frequency compared to the case with full hydrodynamics.

\section{Information on the Supplementary Movies}

The Supplementary Movies provide dynamic visualizations of the structure formation of magnetizable particles subject to an external magnetic field.
We compare the cases including and excluding hydrodynamic coupling for area fractions $\theta \approx 0.15$ (Supplementary Movie 1) and $\theta \approx 0.31$ (Supplementary Movie 2). In both movies, the ``dry'' case without hydrodynamics is depicted on the left, whereas the impact of full hydrodynamic coupling is shown on the right. As in the snapshots in Fig.~1 and~2 in the manuscript, the color scale refers to the magnetization field and the blue arrows indicate emerging flow fields of the carrier liquid.
The duration of both movies covers $20000$ time units.

\newpage

\bibliography{references}